# Crystal growth of hexagonal boron nitride (hBN) from Mg-B-N solvent system under high pressure[#]

N.D. Zhigadlo*

*Laboratory for Solid State Physics, ETH Zurich, Otto-Stern-Weg 1, 8093 Zurich, Switzerland*

**Abstract**

Transparent and colorless hexagonal boron nitride (hBN) single crystals were grown from the Mg-B-N system using high-pressure high-temperature cubic anvil technique. By varying the synthesis conditions we could determine the sequence of phase transformations occurring in the Mg-B-N system, construct the pressure-temperature (*P-T*) phase diagram and discuss the possible growth mechanism. The largest plate-like-shaped hBN crystals with sizes up to 2.5 mm in length and up to 10 μm in thickness were grown at 30 kbar and 1900-2100 °C. The hBN crystals exhibited strong, narrow diffraction peaks typical of well-ordered stacking crystal planes, with the *c*-axis perpendicular to the crystal face. A characteristic Raman peak observed at 1367 cm$^{-1}$ with a full width at half maximum of 8 cm$^{-1}$ corresponds to the $E_{2g}$ vibration mode and indicates the high purity and order of hBN crystals grown by this method. From the practical point of view this work can stimulate further explorations of the Mg-B-N solvent system to obtain isotopically-enriched h$^{10}$BN crystals, which can act as a key element in solid-state neutron detector devices.

*Tel. +41 44 633 2249
 E-mail address: zhigadlo@phys.ethz.ch

## 1. Introduction

Boron nitride (BN) is a synthetic material that has been known for many years (see Ref. [1] and references therein). Due to the specific bonding behavior of boron and nitrogen atoms BN exists in many different structures. The well-known four polymorphs of BN are cubic (cBN), hexagonal (hBN), rhombohedral (rBN) and wurtzite (wBN). The variety of interesting properties of boron-nitrogen materials are closely related to their crystal structures. For example cBN is a hard material, and its mechanical, thermal, and electronic properties are similar to diamond. On the contrary hBN is a soft material, whose softness comes from its layered hexagonal crystal structure, similar to graphite; therefore, hBN is known also as "white graphite". In hBN the atomic planes are built by hexagonal rings of alternating B and N atoms, whose covalent bonds are strong (σ-bonding, $sp^2$-hybridization). By contrast the bonding forces between the atomic planes are weak, being of the van der Waals type (π-bonding) and the layers stack with a somewhat different pattern from graphite. Whereas graphite is an excellent electrical conductor, hBN is a good insulator with a band gap close to 6 eV.

hBN possesses many unique physical and chemical properties such as low density, high melting point, high-temperature stability, high thermal conductivity, low dielectric constant and chemical inertness [1]. All these properties make hBN an interesting material for various electronic and optoelectronic devices, including neutron detectors, ultraviolet light emitters, and deep ultraviolet light detectors [2,3]. In recent years, with the advent of graphene, hBN has been shown to be an excellent gate and substrate, which provides a smooth and flat surface necessary for maximizing carrier mobility in graphene devices [4-6]. In addition to a large range of technical applications, the combination of excellent properties of hBN together with the possibility to assemble it in more complex artificially stacked structures opens a new paradigm in the physics of two-dimensional solids [7].

To achieve the best possible performance in these applications, and to further explore the new physical properties of such two-dimensional systems (complimentary to those of graphene), high-quality hBN single crystals with sufficiently large dimensions are needed. So far the most successful method to obtain single crystals of hBN is the high-pressure and high-temperature growth by using different kinds of solvents [8-10], although some promising results were also obtained at ambient pressure conditions [11-14]. One of the first attempts to grow hBN crystals was carried out by Ishii and Sato [8] which used boron dissolved in a silicon flux under nitrogen atmosphere. After cooling the mixture from a temperature of 1850 °C the authors, for the first time, obtained free standing hBN crystals, measuring 2 mm across and 20 µm in thickness. However, due to carbon impurities, these hBN crystals were yellow in appearance and also contained nitrogen vacancies. Later on, the most successful results in growing hBN crystals were achieved by Watanabe and Taniguchi by using various kinds of solvents (Ni, Ni-Mo, and Ni-Cr) and growth conditions. They showed that at atmospheric pressure, owing to small nitrogen dissolution into the Ni solvent, the growth rates were too low [11,12]. When using Ni-Mo solvent, the grown crystals formed aggregates with a dimension of several hundred micrometers wide and about 10 µm thickness [11]. The use of Ni-Cr solvent was beneficial in increasing the thickness of hBN crystals [13]. The most reliable results were obtained by the use of a reactive Ba-BN solvent under high pressure [9,10]. However, although the Ba-BN system represents a powerful solvent, it is hygroscopic and oxidizes easily in air. Therefore, practical work with such system requires special environmental conditions (dry nitrogen atmosphere), since even slight oxygen contamination may substantially disturb the overall crystal growth process. Thus, one of the challenges in the field is the search for alternative solvent systems which can yield high-quality hBN single crystals.

In this paper, we report on the successful growth of high-quality hexagonal hBN crystals with large sizes by using the Mg-B-N solvent system under high pressure. The optimization of pressure-temperature ($P$-$T$) conditions has been investigated and the possible mechanism for the growth of hBN is discussed.

## 2. Experimental details

For the growth of hBN crystals we used a high-pressure, high-temperature (HPHT) cubic anvil technique, which was developed in our laboratory for growing superconductors and various other compounds [15]. Magnesium flakes (99.99 % pure), amorphous boron powder (99.99 %) and amorphous boron nitride powder (99.9 %) were used as starting materials. A mixture of Mg, B and BN in a molar ratio 1:1.2:0.1 (total mass ~ 1 g) was thoroughly grounded and enclosed in a BN crucible of 8 mm internal diameter and 9 mm length. The benefit of using BN crucibles consisted in minimal system contamination, since crucible and target materials are the same. To remove oxygen impurities and other volatile contaminants before the use, the BN sources were purified by heating them at 2000 °C for 2 h under dynamic vacuum. The BN crucible filled with the Mg-B-BN mixture was placed inside a pyrophylite cube with a graphite heater. Six tungsten carbide anvils generated pressure on the whole assembly. In a typical run, the sample was gradually compressed to the desired pressure at room temperature. While keeping the pressure constant, the temperature was ramped up in 1 h to the desired value, maintained there for 0.5-2 h, and then decreased to 20 °C in 1 h. Afterwards, the pressure was released, and the sample removed. The overall growth procedure lasted for less than 4 h. After mechanically crushing the BN crucible, the high pressure product was heated in vacuum at 750 °C for 0.5 h to remove the excess of Mg. The most suitable conditions for growing sizable bulk hBN crystals were found above 25 kbar in the temperature range comprised between 1800 and 2100 °C. The general morphology and the size of the grown products were monitored by an optical microscope (Leica M 205C). The x-ray quality of the BN crystals and the growth directions were studied at room temperature using a Bruker diffractometer with Cu $K_\alpha$ radiation. To further verify the quality of the grown crystals, Raman spectroscopy was also employed.

## 3. Results and discussion

By varying the synthesis conditions we determined the sequence of phase transformations occurring in the Mg-B-N system. Although the Mg solvent apparently plays an important role, the crystal growth process of hBN is not a simple precipitation from a solution in molten magnesium. Rather, the hBN crystals are the product of a reaction in the ternary Mg-B-N system. We note here that while typically we used a ratio of Mg:B:BN equal to 1:1.2:0.1, one needs to keep in mind that the samples in the real experiment are placed in a BN crucible. At high temperature, the Mg melt and vapor are very aggressive and can react with the BN crucible. Hence we must consider the crucible material as an integral part of the chemical system. It appears, that for the growth of hBN crystals the BN crucible is an ideal choice. Fig. 1 summarizes in a schematical way the resulting *P-T* phase diagram with the majority of the obtained phases. During our exploratory synthesis attempts we observed the formation of black hexagonal prism-shaped crystals, which were involved in the growth process of hBN crystals [16]. Structural x-ray studies revealed that these black hexagonal crystals are a new phase, namely $MgNB_9$ [17]. At low pressures, i.e. below the blue line (Fig. 1), the $MgNB_9$ remained as a stable product. In the vicinity of the phase boundary the $MgNB_9$ phase starts to decompose, with transparent hBN and black-colored $MgB_2$ crystals appearing inside the $MgNB_9$ crystals. After longer soaking times and/or at slightly higher synthesis temperatures and pressures the $MgNB_9$ phase disappears completely and only hBN and $MgB_2$ crystals (attached to each-other) remain in the crucible. Such crystals, which grow by decomposition of $MgNB_9$, most likely serve as seeds for the growth of larger hBN crystals from solution in Mg. The ambient pressure growth of hBN crystals from the Mg-B-N system could be difficult since partial pressure of Mg vapor above molten Mg increases with temperature and at 1800 °C it can be of the order of 50 bar. However, at high pressure the gas phase disappears and only solid and liquid phases remain in the system. In some sense the Mg-B-N system used in the present study is unique, since it allows the simultaneous crystal growth of completely different types of materials: a wide-band semiconductor (hBN) and a superconductor ($MgB_2$). In contrast to the alkaline-earth solvents, the Mg-B-N system has several advantages. Taniguchi and Watanabe [9,10] used a barium boron nitride ($Ba_3B_2N_4$) solvent to dissolve BN powder at 40-50 kbar and 1500-1650 °C. These conditions are very close to cBN/hBN phase equilibrium and result in the appearance of two phases in the final products. Since hBN is the predominant phase at low pressures, our synthetic conditions with synthesis pressure of 30 kbar are more suitable for growing hBN samples. Another significant disadvantage of alkaline-earth based solvents is their high sensitivity to air and moisture, and this clearly avoided in our case.

Based on our systematic investigations we found that at $T \leq 1700$ °C the single crystalline products usually are small in sizes, often not very well crystallizes and form agglomerates. The best conditions which yield a large quantity and the biggest size of hBN crystals are realized at a pressure of 30 kbar and a temperature of 1900-2100 °C. Under these conditions, many thin flaky hBN crystals were found to grow in contact with $MgB_2$ crystals (Fig. 2). The structural and physical properties of $MgB_2$ crystals grown by this process were reported in our previous publications [16]. Here we focus our attention on the observed hBN crystals. It was found that the hBN crystals extended their large planes towards the uniaxial direction of the crucible. This is related to the fact that the graphite furnace has a finite axial extent, inducing a temperature gradient in the high-pressure cell assembly (for details see Ref. [18]). The hBN crystals obtained by this method were colorless and transparent and had platelet shapes with smooth surfaces. The largest crystals were up to 2.5 mm in in-plane dimension and up to 10 µm in thickness (Fig. 3). The grown crystals were often fragmented into several pieces and they could easily be separated by applying a small mechanical force.

The hBN structure was confirmed by x-ray diffraction to have a lattice constant $c$ of 6.66 Å. All grown hBN crystals exhibited strong, narrow peaks for the (002), (004), and (006) planes, corresponding to stacked planes in the $c$-direction (Fig. 4). The full width at half maximum (FWHM) for the (006) peak is 0.15°. This value is considerably smaller than that reported in Ref. [14] and confirms that crystals consist of well-ordered stacking crystal planes.

The Raman spectrum of hBN crystal is presented in Fig. 5. Only one Raman line with FWHM of about 8 cm$^{-1}$ at 1367 cm$^{-1}$ is observed. This peak corresponds to the $E_{2g}$ in-plane vibration mode of hBN [19] and is analogous to the well-known G peak in graphene [20]. Other lines, such as those at 1056 cm$^{-1}$ and 1306 cm$^{-1}$ of cBN, or 1580 cm$^{-1}$ of carbon were not detected in the explored range (1000 cm$^{-1}$ to 1800 cm$^{-1}$). This indicates that our method utilizes conditions suitable for the growth of hBN and that there are neither inclusions of other impurities nor the formation of metastable cubic or wurtzitic BN phases. The FWHM value we obtain is one of the narrowest reported to date for hBN crystals [9] and indicates a well-ordered crystal with minimal defects. All together our results indicate that the grown hBN crystals are of high crystal quality and uniformity.

Another benefit of using a precursor containing Mg, B, and BN is the possibility to grow $^{10}$B-enriched hBN crystals. This is of particular interest if one considers the potential of hBN crystals as solid-state neutron detectors [21-23]. Typically these detectors work on the principle of converting thermal neutrons into alpha-particles, which can then be sensed by a p-n junction. Thermal neutrons have a low probability of interacting with conventional semiconductor materials. However, by using a semiconductor containing $^{10}$B material, such as h$^{10}$BN, with a large thermal neutron cross-section, this problem can easily be overcome. So far, hBN crystal growth has relied mostly on cBN [10] or BN [13] powder source materials which contain boron in its natural abudance, i.e. 20 % $^{10}$B and 80 % $^{11}$B. The advantage of our starting composition (1Mg + 1.2B + 0.1BN) consist in the use of large amounts of amorphous boron powder, which is commercially available in enriched $^{10}$B form, whereas this is not the case for BN powder. Thus, we conclude that the success of growth method presented here opens the door to the growth of isotopically-enriched h$^{10}$BN single crystals, which in the future can be used in solid-state neutron detector devices.

## 4. Conclusions

Single crystals of hBN with a size up to 2.5 mm in length and up to 10 μm in thickness have been grown at high pressure and high temperature from a precursor containing Mg, B, and BN. The crystal growth process is very peculiar and involves an intermediate nitride, namely MgNB$_9$, which seems to play an important role in the growth of hBN. The x-ray data confirmed the hexagonal crystal structure of hBN. The Raman peak observed at 1367 cm$^{-1}$ with an FWHM of 8.0 cm$^{-1}$, corresponds to the $E_{2g}$ vibration mode for hBN and indicates the high purity and order of the single crystals grown by this process. The results obtained in the present work may be viewed as an additional confirmation of the effectiveness of high pressure method in the growth of hBN crystals.

**Acknowledgements**

The author would like to thank his colleagues J. Jun, S. Kazakov, and J. Karpinski for their support, helpful assistance and valuable discussions. He is also grateful to M.

LeTacon for measuring the Raman spectra and T. Shiroka for useful comments. This research was partially supported by the Swiss National Science Foundation, the National Center of Competence in Research MaNEP (Materials with Novel Electronic Properties) program.

# References

[1] R. Haubner, M. Wilhelm, R. Weissenbacher, B. Lux, Boron nitrides-properties, synthesis and applications, In: Structure and Bonding, vol. 102, Springer-Verlag Berlin Heidelberg (2002) 1-45.

[2] K. Watanabe, T. Taniguchi, T. Niiyama, K. Miya, M. Taniguchi, Nature Photonics 3 (2009) 591-594.

[3] K. Watanabe, T. Taniguchi, Int. J. Appl. Technol. 8, 5 (2011) 977-989.

[4] C. R. Dean, A. F. Young, I. Meric, C. Lee, L. Wang, S. Sorgenfrei, K. Watanabe, T. Taniguchi, P. Kim, K. L. Shepard, J. Hone, Nature Nanotechnology 5 (2010) 722-726.

[5] W. Yang, G. Chen, Z. Shi, C.-C. Liu, L. Zhang, G. Xie, M. Cheng, D. Wang, R. Yang, D. Shi, K. Watanabe, T. Taniguchi, Y. Yao, Y. Zhang, G. Zhang, Nature Materials 12 (2013) 792-797.

[6] W. Gannett, W. Regan, K. Watanabe, T. Taniguchi, M. F. Crommie, A. Zettl, Appl. Phys. Lett. 98 (2011) 1-3.

[7] S. Dai, Z. Fei, Q. Ma, A. S. Rodin, M. Wagner, A. S. McLeod, M. K. Liu, W. Gannett, W. Regan, K. Watanabe, T. Taniguchi, M. Thiemens, G. Dominguez, A. H. Castro Neto, A. Zettl, F. Keilmann, P. Jarillo-Herrero, M. M. Fogler, and D. N. Basov, Science 343 (2014) 1125-1129.

[8] T. Ishii, T. Sato, J. Cryst. Growth 61 (1983) 689-690.

[9] K. Watanabe, T. Taniguchi, H. Kanda, Nature Materials 3 (2004) 404-409.

[10] T. Taniguchi, K. Watanabe, J. Cryst. Growth 303 (2007) 525-529.

[11] Y. Kubota, K. Watanabe, O. Tsuda, T. Taniguchi, Science 317 (2007) 932-934.

[12] Y. Kubota, K. Watanabe, T. Taniguchi, Jpn. J. Appl. Phys. 46 (2007) 311-314.

[13] Y. Kubota, K. Watanabe, O. Tsuda, T. Taniguchi, Chem. Mat. 20, 5 (2008) 1661-1663.

[14] T. B. Hoffman, B. Clubine, Y. Zhang, K. Snow, J. H. Edgar, J. Cryst. Growth 393 (2014) 114-118.

[15] N. D. Zhigadlo, S. Katrych, Z. Bukowski, S. Weyeneth, R. Puzniak, J. Karpinski, J. Phys.: Condens. Matter 20 (2008) 342202; R. T. Gordon, N. D. Zhigadlo, S. Weyeneth,

S. Katrych, R. Prozorov, Phys. Rev. B 87 (2013) 094520; N. D. Zhigadlo, J. Cryst. Growth 382 (2013) 75-79; N. D. Zhigadlo, J. Cryst. Growth 395 (2014) 1-4.

[16] J. Karpinski, M. Angst, J. Jun, S. M. Kazakov, R. Puzniak, A. Wisniewski, J. Roos, H. Keller, A. Perucchi, L. Degiorgi, M. R. Eskildsen, P. Bordet, L. Vinnikov, A. Mironov, Supercond. Sci. Technol. 16 (2003) 221; J. Karpinski, N.D. Zhigadlo, S. Katrych, R. Puzniak, K. Rogacki, R. Gonnelli, Phys. C 456 (2007) 3-13.

[17] A. Mironov, S. Kazakov, J. Jun, J. Karpinski, Acta Cryst. C58 (2002) i95-i97.

[18] N. D. Zhigadlo, S. Weyeneth, S. Katrych, P. J. W. Moll, K. Rogacki, S. Bosma, R. Puzniak, J. Karpinski, and B. Batlogg, Phys. Rev. B 86 (2012) 214509.

[19] R. Geick, C. H. Perry, G. Rupprecht, Phys. Rev. 146 (1966) 543-547.

[20] R. Arenal, A. C. Ferrari, S. Reich, L. Wirtz, J. V. Mevellec, S. Lefrant, A. Rubio, A. Loiseau, Nano Lett. 6 (2006) 1812-1816.

[21] J. Uher, S. Pospisil, V. Linhart, M. Schieber, Appl. Phys. Lett. 90 (2007) 124101.

[22] J. Li, R. Dahal, S. Majety, J. Y. Lin, H. X. Jiang, Nucl. Instrum. Meth. A 654 (2011) 417-420.

[23] S. Barman, Europhys. Lett. 96 (2011) 16004.

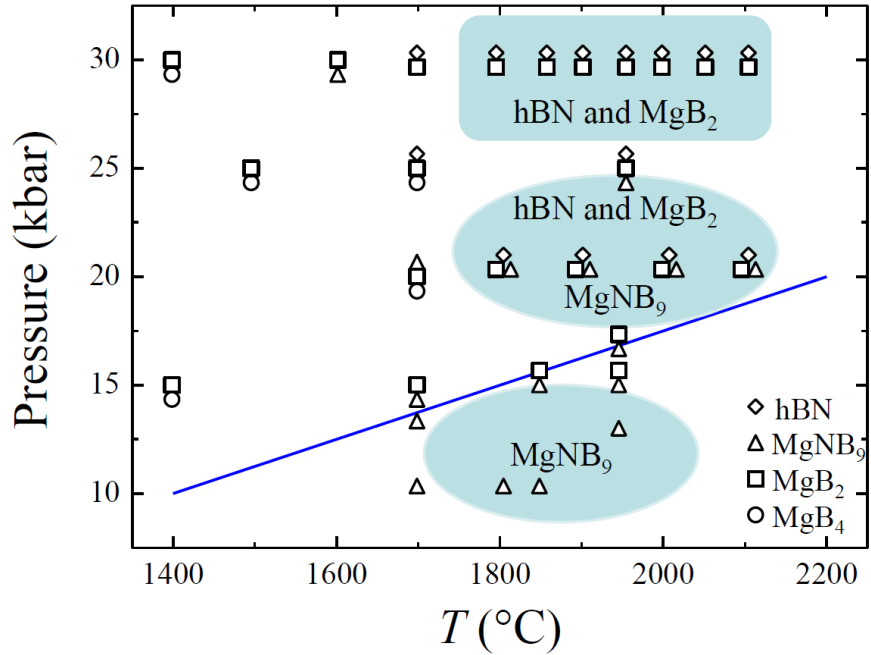

**Fig. 1.** Pressure-temperature phase diagram of Mg-B-N system plotted using our experimental data and some data from Ref. [16]. Symbols show the phases observed in the samples. While $MgB_4$ appeared only in some samples, Mg was present in all of them, but it could be removed by heating the as-grown samples in vacuum at 750 °C. At $T \leq$ 1700 °C the single crystalline products are usually small in size, not well crystallized and form agglomerates. Highlighted areas show the main bulk phases observed in the high-pressure products synthesized at $T \geq 1800$ °C. The solid blue line represented the boundary of the phase stability of $MgNB_9$. Above this line the $MgNB_9$ compound appears only as a metastable secondary phase. The most suitable conditions for growing hBN (together with $MgB_2$) crystals are found above 25 kbar in the temperature range comprised between 1800 and 2100 °C.

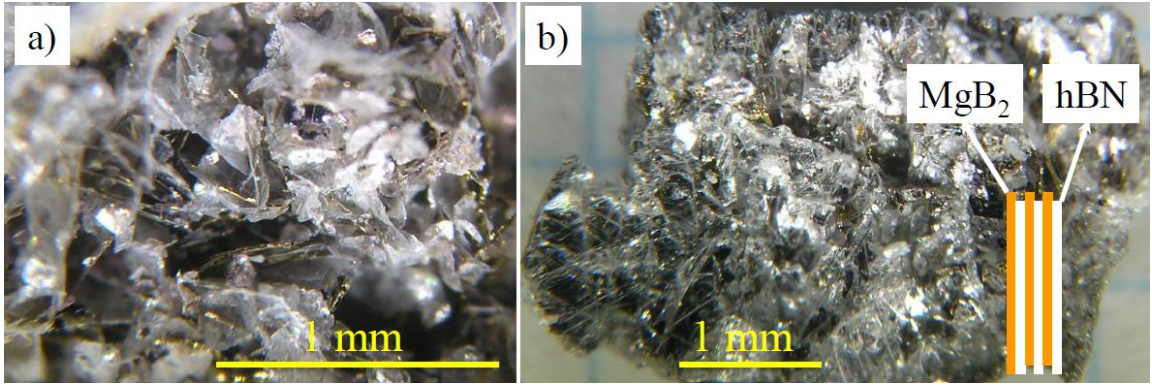

**Fig. 2.** Optical images of the top-down perspective (a) and the edge-on view (b) of the high pressure products after removing the excess of Mg through vacuum annealing at 750 °C. Two main crystalline phases are visible: transparent and colorless hBN single crystals and black-colored $MgB_2$ single crystals. The schematic drawing in b) suggests the possible mechanism of hBN + $MgB_2$ crystal growth by decomposition of the intermediate nitride compound $MgNB_9$.

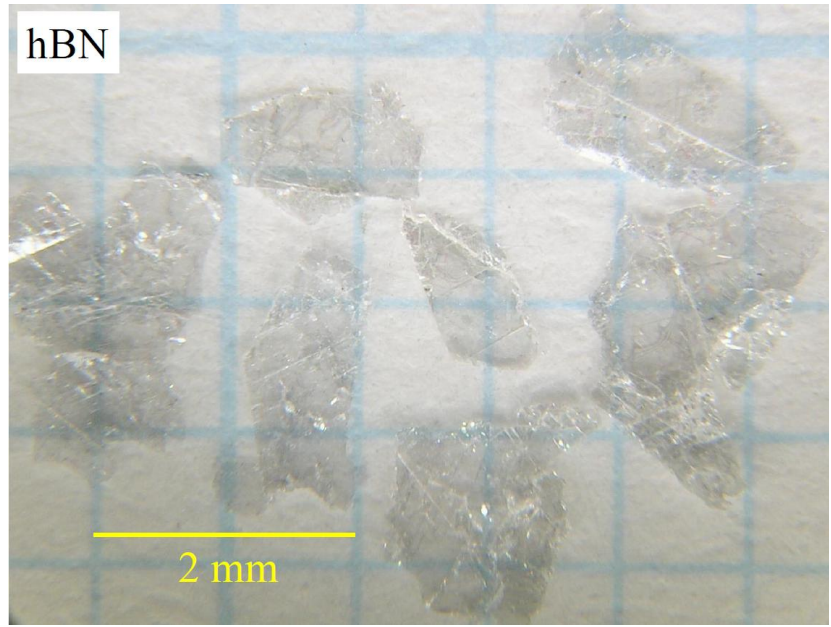

**Fig. 3.** Optical image of hexagonal boron nitride (hBN) crystals with various shapes and sizes, as extracted mechanically from the high-pressure products grown at 30 kbar and at different temperatures (1900-2100 °C).

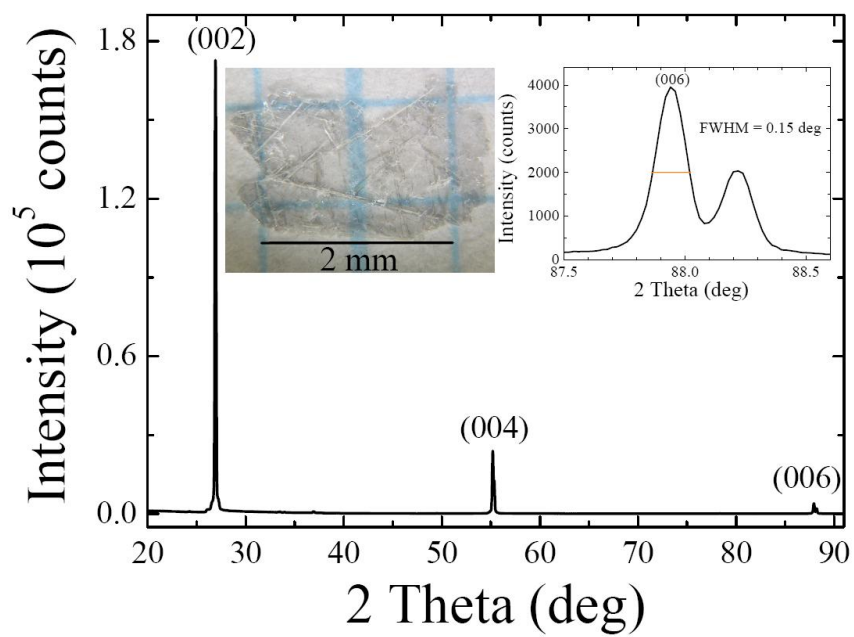

**Fig. 4.** Spectrum of x-ray diffraction of a hBN single crystal (left inset) grown at 30 kbar and 2000 °C. The known crystallographic planes of hBN are labeled above the corresponding peaks. The right inset shows a magnified version of the (006) peak (FWHM = 0.15 °).

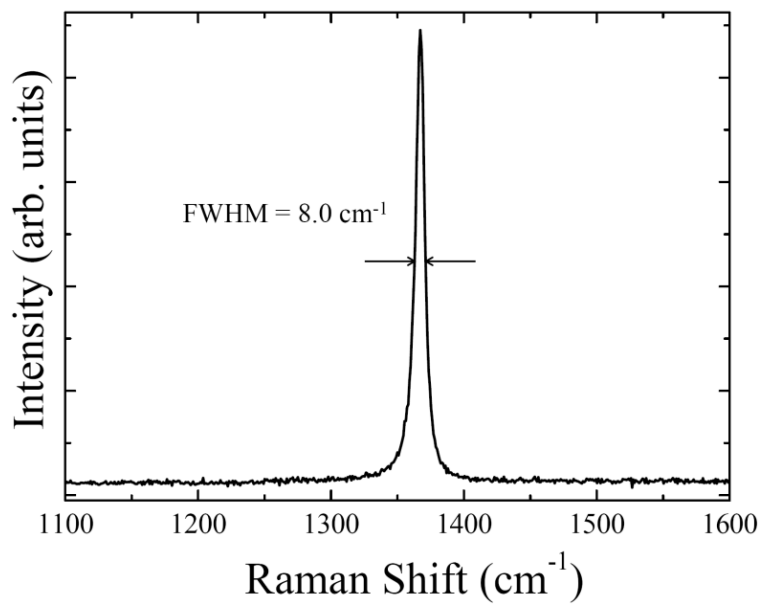

**Fig. 5.** Raman spectrum of a hBN single crystal grown at high pressure. The peak at 1367 cm$^{-1}$ with FWHM of 8.0 cm$^{-1}$ corresponds to the $E_{2g}$ vibration mode of hBN and is indicative of the high purity and order of the single crystals grown by this process.